\documentclass[twocolumn,showpacs,preprintnumbers,amsmath,amssymb,superscriptaddress]{revtex4}
\usepackage{graphicx}
\usepackage{dcolumn}
\usepackage{bm}
\usepackage{soul}
\usepackage{color}
\usepackage{epstopdf}
\usepackage[version=3]{mhchem}
\usepackage{lipsum}
\usepackage[outercaption]{sidecap}
\usepackage{floatrow}
\begin{document}

\title{Near-infrared photothermal response of plasmonic gold-coated nanoparticles in tissues}
\author{Vu T. T. Duong}
\affiliation{Institute of Physics, Vietnam Academy of Science and Technology, 10 Dao Tan, Hanoi 10000, Vietnam}
\email{vuthuyduong80@gmail.com}
\author{Anh D. Phan}
\affiliation{Institute of Physics, Vietnam Academy of Science and Technology, 10 Dao Tan, Hanoi 10000, Vietnam}
\affiliation{Department of Physics, University of Illinois, 1110 West Green St, Urbana, Illinois 61801, USA}
\email{adphan35@gmail.com}
\author{Nghiem T. H. Lien}
\affiliation{Institute of Physics, Vietnam Academy of Science and Technology, 10 Dao Tan, Hanoi 10000, Vietnam}
\author{Do T. Hue}
\affiliation{Institute of Physics, Vietnam Academy of Science and Technology, 10 Dao Tan, Hanoi 10000, Vietnam}
\author{Do Q. Hoa}
\affiliation{Institute of Physics, Vietnam Academy of Science and Technology, 10 Dao Tan, Hanoi 10000, Vietnam}
\author{Do T. Nga}
\affiliation{Institute of Physics, Vietnam Academy of Science and Technology, 10 Dao Tan, Hanoi 10000, Vietnam}
\email{dtnga1711@gmail.com}
\author{Tran H. Nhung}
\affiliation{Institute of Physics, Vietnam Academy of Science and Technology, 10 Dao Tan, Hanoi 10000, Vietnam}
\author{N. A. Viet}
\affiliation{Institute of Physics, Vietnam Academy of Science and Technology, 10 Dao Tan, Hanoi 10000, Vietnam}
\date{\today}

\begin{abstract}
We propose a new approach to understand the time-dependent temperature increasing process of gold-silica core-shell nanoparticles injected into chicken tissues under near-infrared laser irradiation. Gold nanoshells strongly absorb near-infrared radiations and efficiently transform absorbed energy into heat. Temperature rise given by experiments and numerical calculations based on bioheat transfer are in good agreement. Our work improves the analysis of a recent study [Richardson et al., Nano Lett. {\bf 9}, 1139 (2009)] by including effects of the medium perfusion on temperature increase.
The theoretical analysis can also be used to estimate the distribution of nanoparticles in experimental samples and provide a relative accuracy prediction for the temperature profile of new systems. This methodology would provide a novel and reliable tool for speeding up photothermal investigations and designing state-of-the-art photothermal devices.
\end{abstract}

\pacs{}
\maketitle
\section{Introduction}

Tremendous efforts have been dedicated to photothermal effect due to a variety of applications such as selectively destroying cancer cells \cite{5,6,7}, delivering drugs \cite{8,9} and improving the quality of magnetic resonance imaging \cite{10}. The photothermal therapy replies upon the quantum behavior of nanomaterials, particularly metallic nanoparticles, under the illumination of near-infrared (NIR) laser light. Strong interactions between the nanoparticles and incident light result from the excitation of the collective oscillations of conduction electrons-surface plasmon resonances \cite{11,12} - within the particles. By exciting these surface plasmon resonances, metallic nanoparticle samples can be locally heated to a few hundred degree Celsiuses when irradiated by relatively low-intensity light \cite{17}. It has been shown that cancerous cells can be destroyed at temperatures above 43 $^0$C \cite{18,19} due to irreversible protein denaturation. To reduce the photobleaching and photodamage of organisms, NIR light sources are extremely useful as tools for photothermal therapy. NIR light is absorbed less and penetrates the tissues without much damage. As a result, disease areas can be locally annihilated without impacting healthy cells. The laser wavelengths employed in photothermal treatment range from 650 to 900 nm \cite{13,14}. In this range, the biological transparency window provides tissue penetration depths of approximately 10 cm \cite{15,16}.

Properly synthesizing the size of nanostructures can modulate the plasmonic wavelength for the NIR region to maximize the absorption of the light energy by gold nanostructures while effectively releasing the heat. The development of nanofabrication technology has enabled the design of nanostructures with desired shapes and sizes. Among them, gold nanoshells have been extensively explored as the well-known NIR photothermal conversion agents. 

Theoretical models have been introduced to describe plasmonic photothermal light-to-heat conversion and how important factors affect sample heating. In Ref.\cite{2,3,33,34}, authors showed the temperature profile outside nanoparticles in suspension using the heat diffusion conditions at interfaces. This approach successfully explains the qualitative dependence of the steady state temperature $T_{max}$ on the excitation laser light and absorbance of the nanostructures but it fails to quantitatively estimate sample temperature since it considers the temperature distribution of a single nanoparticle. In this context, Govorov and his co-workers \cite{3} described the laser-induced heating profile by employing the thermal transfer equation and an equivalent solution of Poisson equation from electrostatics. Govorov's model captures the collective effects of nanoparticles on temperature increase to be more realistic than the prior thermal single particle model in water droplet where the perfusion effects play a minor role in the heat transfer process. Another model \cite{3,4} based on the energy balance equation in homogeneous and open systems excellently provides analytical fitting functions $T(t)=T_0 + \frac{A}{B}(1-e^{-Bt})$ and $T(t)=T_0+(T_{max}-T_0)e^{-Bt}$ for time-dependent temperature as turning on and off the laser excitation, respectively, where $T_0$ is the ambient temperature, $A$ and $B$ are the fitting parameters which capture physical properties of the system. The theoretical analysis associated with experimental data has been used to determine the photothermal efficiency but it is hard to predict reliably unknown experimental systems. In this work, we incorporate effects of medium perfusion on bioheat transfer into the standard thermal diffusion equation of Govorov's model to develop a simple model for investigating the photothermal effect of silica-gold core-shell nanoparticles injected into chicken tissues where the perfusion effects cannot be ignored. Our theoretical calculations are in very good agreement with experimental measurements of the temperature profile at different intensities of laser illumination. This method can provide good quantitative predictions for the time-dependent temperature of unexplored experiments.

\section{Experimental Section}
The nanocomposites composed of a 128$\pm$10 nm $SiO_2$ core and a 16$\pm$ 3 nm thick gold shell were dispersed in water solutions which were synthesized in the same process with Ref.\cite{1}. Their UV-visible absorption spectra at room temperature were recorded by using a JASCO-V570-UV-Vis-NIR spectrometer. Their morphology was studied by using a transmission electron microscope (TEM, JEM1011). We injected the cores-shell nanoparticles into chicken tissues. Experimental samples (of the chicken tissues injected Au nanostructures) with dimensions of 4$\times$4$\times$4 $mm^3$ were irradiated by a NIR-diode laser (808 nm wavelength) with intensities of 2, 4 and 6 W/$cm^2$. The laser beam diameter is about 1 mm. Just after illumination with the laser, the temperature is determined by the resistance change of the PT 100-CRZ sensor (Hayashi Denko, Japan) placed under the chicken tissue sample for 7 minutes. The temperature sensor PT 100 was connected with a data acquisition device NI USB-6058 (National Instruments) controlled by a computer. The experiments were performed at room temperature 22$^0$ C. 


\section{Theoretical background of the photothermal effect}

The original Mie theory has widely and successfully given the absorption and scattering spectra of a spherical nanoparticle embedded in a homogeneous and isotropic medium \cite{20}. The approach shows a good agreement with experimental results. The Mie theory has been recently developed to investigate optical properties of large nanoparticles and composite systems. Applying the Mie approach to the core-shell nanospheres provides \cite{22}
\begin{eqnarray}
Q_{ext} &=& -\frac{2\pi}{k_m^2}\sum_{n=1}^{\infty}(2n+1){Re}\left(a_n + b_n \right), \nonumber\\
Q_{scat} &=& \frac{2\pi}{k_m^2}\sum_{n=1}^{\infty}(2n+1)\left(\left | a_n \right|^2  + \left | b_n \right |^2 \right),\nonumber\\
Q_{abs} &=& Q_{ext}  - Q_{scat},
\label{eq:1}
\end{eqnarray}
where 
\begin{eqnarray}
a_n=-\frac{U_n^{TM}}{U_n^{TM}+iV_n^{TM}} ,\quad b_n=-\frac{U_n^{TE}}{U_n^{TE}+iV_n^{TE}},
\end{eqnarray}

\begin{eqnarray}
U_n^{TM} = \begin{vmatrix} j_n(k_cR_c) & j_n(k_sR_c) & y_n(k_sR_c) & 0 \\ \cfrac{\Psi_n^{'}(k_cR_c)}{\varepsilon_c} & \cfrac{\Psi_n^{'}(k_sR_c)}{\varepsilon_s} & \cfrac{\Phi_n^{'}(k_sR_c)}{\varepsilon_s}& 0 \\ 
0 & j_n(k_sR_s) & y_n(k_sR_s) & j_n(k_mR_s) \\
0 & \cfrac{\Psi_n^{'}(k_sR_s)}{\varepsilon_c} & \cfrac{\Phi_n^{'}(k_sR_s)}{\varepsilon_s} & \cfrac{\Psi_n^{'}(k_mR_s)}{\varepsilon_m}
\end{vmatrix}, \nonumber\\
V_n^{TM} = \begin{vmatrix} j_n(k_cR_c) & j_n(k_sR_c) & y_n(k_sR_c) & 0 \\ \cfrac{\Psi_n^{'}(k_cR_c)}{\varepsilon_c} & \cfrac{\Psi_n^{'}(k_sR_c)}{\varepsilon_s} & \cfrac{\Phi_n^{'}(k_sR_c)}{\varepsilon_s}& 0 \\ 
0 & j_n(k_sR_s) & y_n(k_sR_s) & y_n(k_mR_s) \\
0 & \cfrac{\Psi_n^{'}(k_sR_s)}{\varepsilon_c} & \cfrac{\Phi_n^{'}(k_sR_s)}{\varepsilon_s} & \cfrac{\Phi_n^{'}(k_mR_s)}{\varepsilon_m}
\end{vmatrix},\nonumber\\
\label{eq:2}
\end{eqnarray}
where $R_c$ and $R_s$ are the inner and outer radius of the core-shell nanostructure, respectively, $V_n$ and $U_n$ are determinants, $j_n(x)$ is the spherical Bessel function of the first kind, $y_n(x)$ is the spherical Neumann function, and $\Psi(x)=xj_n(x)$ and $\xi_n(x) = xy_n(x)$ are the Riccati-Bessel functions. $U_n^{TE}$ and $V_n^{TE}$ are obtained by replacing the dielectric function in Eq.(\ref{eq:2}) with the permeability. The dielectric functions of core, shell and surrounding (water) medium of the core-shell nanoparticles are $\varepsilon_c$, $\varepsilon_s$ and $\varepsilon_m \approx 1.77$, respectively. The wavenumber is $k_i=2\pi\sqrt{\varepsilon_i}/\lambda$ with $i=s, c$, and $m$. $\lambda$ is the wavelength of incident light in vacuum. In our calculations, the dielectric function of silica ($\varepsilon_c$) is taken from Ref.\cite{23}, while the dielectric function of gold ($\varepsilon_s$) is described by the Lorentz-Drude model with several oscillators \cite{24}
\begin{eqnarray}
\varepsilon_s = 1- \frac{f\omega_p^2}{\omega^2-i\omega\Gamma_0} + \sum_{j=1}^{5}\frac{f_j\omega_p^2}{\omega_j^2-\omega^2+i\omega\Gamma_j},
\end{eqnarray}
where $f_0$ and $f_j$ are the oscillator strengths, $\omega_p$ is the plasma frequency for gold, and $\Gamma_0$ and $\Gamma_j$ are the damping parameters. All parameters in this model come from Ref. \cite{24}. However, due to the fact that the shell thickness is 16 nm, the finite size effect becomes important. This effect can be added to the model by modifying the parameter $\Gamma_0 \equiv \Gamma_0+Bv_F/(R_s-R_c)$, where $v_F$ is the gold Fermi velocity, and $B$ is the parameter characterizing the scattering processes. Values of B have been chosen between 0.1 and 5 \cite{35,36,37}. Modifying the bulk scattering rate reproduces the relation among the width of the resonance peak at 830 nm, the shape of the shoulder at around 650 nm and gold layer thickness. Choosing $B = 1.5$ gives a reasonable fit between the theoretical calculations and experiments.

After injecting 2 $\mu l$ of gold nanoshells solution (concentration of $ 10^{11}$ particles/ml) into chicken tissue samples, the samples were illuminated by a laser with a wavelength of 808 nm in the region of the therapeutic window. The laser light, thus, can go through, touch the NPs and excite the surface localized resonance. The nanostructures were heated and effectively converted to heat energy. Suppose that the efficiency of the light-to-heat conversion is 100 $\%$ and the particles are randomly dispersed in a spherical region of radius $R$, the heat source density due to absorbed energy on nanoparticles is $A = NQ_{abs}I_0$, where $N$ is the number of particles per unit volume in the chicken tissue and $I_0$ is the illuminating intensity. The particle density $N$ is the number of injected nanoshells divided by the volume of the chicken tissue. Here we also assume that the particle concentration is homogeneous. Temperature variation of experimental samples $\Delta T(r,t)$ induced by the photothermal effect of gold materials is theoretically described by the Pennes bioheat transfer equation \cite{25,40,26} in spherical coordinates 
\begin{eqnarray}
\frac{1}{\kappa}\frac{\partial \Delta T}{\partial t} = \frac{1}{r^2}\frac{\partial}{\partial r}\left(r^2\frac{\partial \Delta T}{\partial r} \right) - \frac{\Delta T}{\kappa\tau} + \frac{A}{k},
\label{eq:3}
\end{eqnarray}
where $k$ and $\kappa=k/(\rho c)$ are the thermal conductivity and thermal diffusivity of the tissue, respectively,  $\rho$ is the mass density, $c$ is  the specific heat, and $\tau$ is the perfusion time constant. Different materials and media have different values of $\tau$, so the value used in the theoretical calculations is taken from previous studies \cite{28,29,30,31,32}. The presence of the heat generation $A$ only exists within the localized domain of the nanoshells because there are no nanoparticles outside the region. This assumption, thus, suggests rewriting Eq.(\ref{eq:3}) as
\begin{eqnarray}
\frac{1}{\kappa}\frac{\partial \Delta T}{\partial t} &=& \frac{1}{r^2}\frac{\partial}{\partial r}\left(r^2\frac{\partial \Delta T}{\partial r} \right) - \frac{\Delta T}{\kappa\tau} + \frac{A}{k}, \quad 0 \leq r \leq R, \nonumber\\
\frac{1}{\kappa}\frac{\partial \Delta T}{\partial t} &=& \frac{1}{r^2}\frac{\partial}{\partial r}\left(r^2\frac{\partial \Delta T}{\partial r} \right) - \frac{\Delta T}{\kappa\tau}, \qquad R \leq r.
\label{eq:4}
\end{eqnarray}

Solving Eq.(\ref{eq:4}) associated with the boundary conditions $\Delta T(R^-,t)= \Delta  T(R^+,t)$ and $\Delta  T'(R^-,t)= \Delta  T'(R^+,t)$, and the finite value of temperature change at $r=0$ and $\infty$ gives temporal and spatial temperature distributions in the spherical region. The temperature at the center of the localized spherical domain of the gold nanoshells is used to determine temperature measurement using thermal probes \cite{25}. The temperature at $r = 0$ is

\begin{eqnarray}
\Delta  T(r=0,t)&=&\frac{A}{k}\left[ -\kappa\int_0^t e^{-t'/\tau}\mbox{erfc}\left(\frac{R}{2\sqrt{\kappa t'}} \right)dt' \right. \nonumber\\
&-& \left. R\int_0^t e^{-t'/\tau}\sqrt{\frac{\kappa}{\pi t'}}\exp\left(-\frac{R^2}{4\kappa t'}\right)dt' \right.\nonumber\\
&+& \left. \kappa\tau(1-e^{-t/\tau}) \right].
\label{eq:5}
\end{eqnarray}

\section{Results and discussions}

\begin{figure}[htp]
\includegraphics[width=8cm]{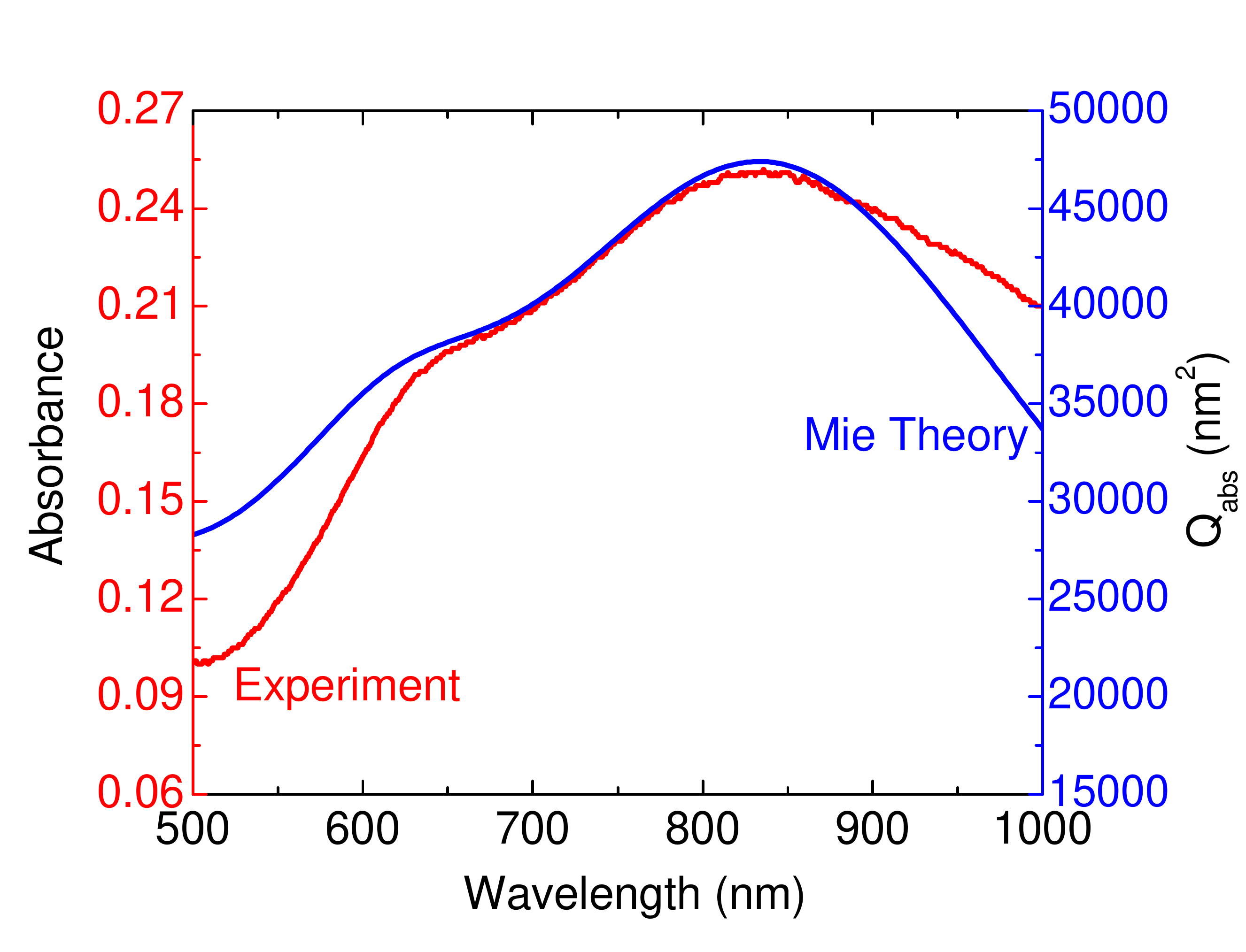}
\caption{\label{fig:1}(Color online) Theoretical (blue) and experimental (red) UV-vis absorption spectrum obtained from aqueous suspensions of Au nanoshells with inner and outer radius of 64 amd 80 nm. }
\end{figure}

Figure \ref{fig:1} shows the experimental and theoretical ultraviolet-visible absorption spectra of the gold nanoshell solution having an absorbance resonance peak near 836 nm. There is an excellent agreement between the absorption spectra given by experiment and Mie theory for nanostructures with a 128 nm silica core and a shell thickness of 16 nm. The discrepancy between our Mie theory calculation and experiments is possibly due to partial particle aggregation. The optical peak at $\lambda \approx 836$ nm is attributed to dipolar localized surface plasmon resonance while the quadrupolar mode is responsible for the shoulder near 640 nm. For larger nanoshells, the absorption feature becomes greater. The absorbed light energy contributes to the heating process of the surrounding environment, leading to an increase in temperature of the samples. 

After gold nanoshell injection, the nanoparticles are assumed to be uniformly distributed in the tissue. The hypothetical radius $R$ is considered as an effective radius which gives an equality between the volume of the spherical region and that of the chicken tissue.

\begin{figure}[htp]
\includegraphics[width=8cm]{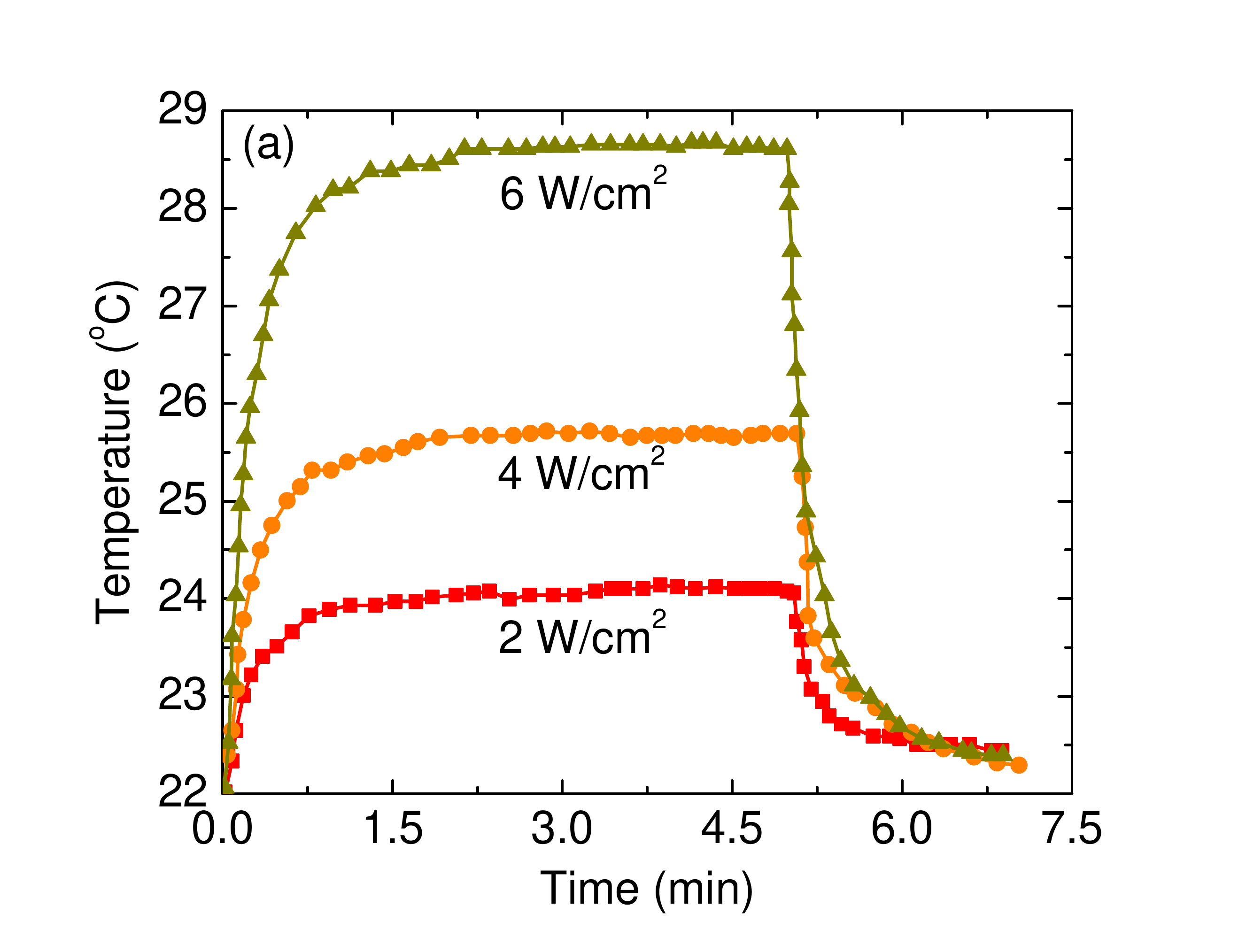}\\
\includegraphics[width=8cm]{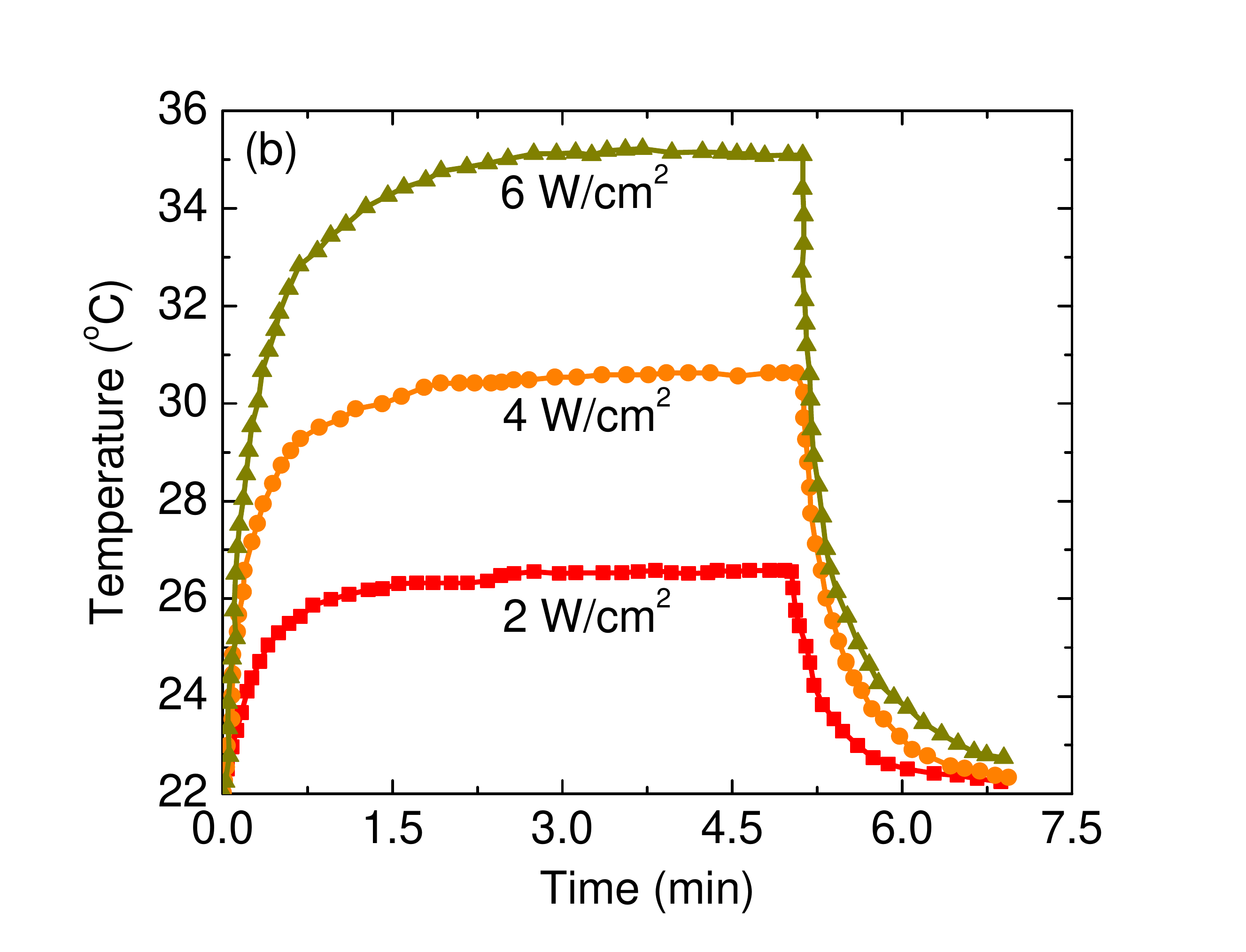}
\caption{\label{fig:2}(Color online) Photothermal heating curves of chicken tissues in the (a) absence and (b) presence of gold nanoshells under NIR illumination at various laser intensities. }
\end{figure}

Time-dependent temperature of chicken tissue samples with and without core-shell nanostructures is shown in Fig. \ref{fig:2}. In our measurement, experiments were carried out at approximately $22$ $^0$C. The suspension reaches the steady-state temperature after around 1 minute of excitation, while turning off the light leads to the significant decrease of temperature from maximum temperature to room temperature. In the absence of gold materials, laser exposure causes an increase of temperature in chicken tissues. The observation has also been reported in many previous studies \cite{27,21}. Heat induced by the laser source raises not only the environment temperature near the experiment set-up, but also the samples. The perturbation is unpreventable. Interestingly, the temperature variation of the chicken tissue under laser irradiation with $I_0 = 2$ W/$cm^2$ behaves qualitatively and quantitatively in the same manner as a prior paper executing the analogous process with distilled water \cite{21}. Besides, water covers 90 $\%$ of the chicken tissues. These findings suggest that physical quantities of our chicken tissue samples are very similar to that of water $k \approx 0.6$ W/m/K, $\kappa \approx 1.43\times 10^{-7}$ $m^2$/s. The presence of gold nanoshells generates more heat in experimental samples due to pronounced effect of plasmonic heating. 

\begin{figure}[htp]
\includegraphics[width=8cm]{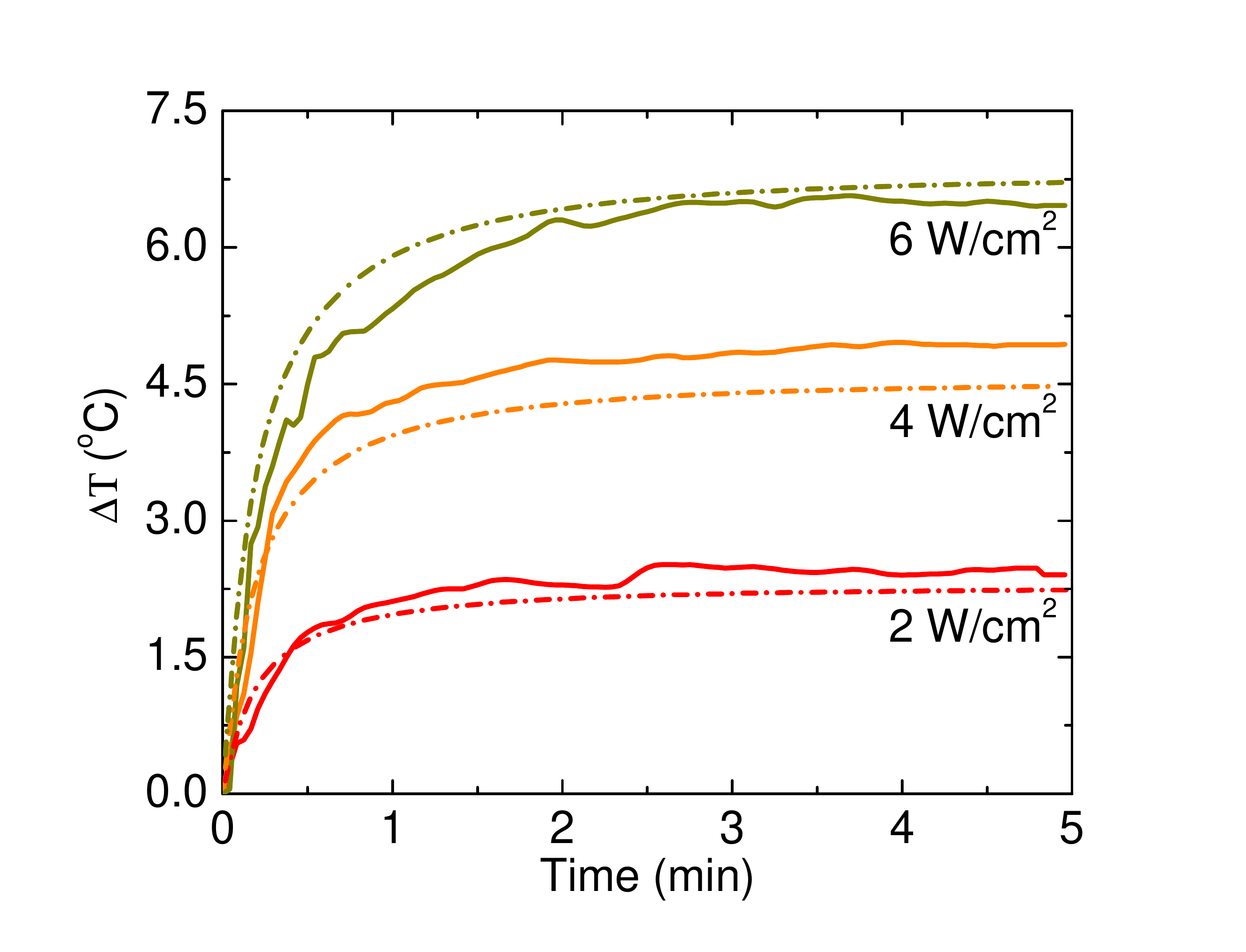}
\caption{\label{fig:3}(Color online) Temperature rise due to absorbed energy of gold nanoshells in samples of chicken tissue as a function of time at various laser intensities. The dashed-dotted lines corresponds to theoretical calculations in the same conditions with equivalent measurements and $\tau = 200$ s \cite{28,29,30,31,32}.}
\end{figure}

Figure \ref{fig:3} presents the photothermal effect of gold nanoshells on the temperature increase as a function of time at $I_0 = 2$, 4, and 6 W/$cm^2$. We suppose that the temperature-increasing process caused by  surface plasmon resonance absorption of silica-gold nanostructures is completely independent of the heating of the chicken tissues without the gold nanoshells. Thus, $\Delta T$ theoretically calculated by Eq.(\ref{eq:5}) can be experimentally estimated by subtracting the experimental data in Fig. \ref{fig:2}a from \ref{fig:2}b. Numerical results closely agree with the experimental measurements.

According to Ref.\cite{3}, the steady-state temperature increase can be estimated using
\begin{eqnarray}
\Delta  T &=&2\Delta  T_{max}R_sNA_{beam}\ln\left[\frac{l_{opt}}{R_{beam}}\right], \nonumber\\
\Delta  T_{max} &=& \frac{Q_{abs}I_0}{4\pi k R_s},
\label{eq:6}
\end{eqnarray}
where $\Delta  T_{max}$ is the maximum temperature at the nanoparticle surface, $l_{opt}$ is the optical path length, $A_{beam}\approx 0.785$ $mm^2$ and $R_{beam} \approx 0.5$ $mm$ are the light spot area and the beam radius, respectively. Other parameters are $Q_{abs} \approx 47400$ $nm^2$, $R_s \approx 72.5$ $nm$. For experiments in chicken tissue, the absorbance is approximately $0.015$, the molar extinction coefficient is $7.53\times 10^{10}$ $M^{-1}cm^{-1}$, and the optical path length is $l_{opt} \approx 0.4$ $cm$. Substituting all the parameters into Eq.(\ref{eq:6}), the temperature changes are: $\Delta T \approx 1.4$, $2.8$ and $4.2$ $^0C$ for $I_0 =2$, $4$, and $6$ $W/cm^2$, respectively. It is clear that our model gives better description for experimental data than the prior work. 

There are three fundamental differences between our approach and Govorov's model in Ref. \cite{3}. First, Govorov's model considers the heat source due to nanostructures in the region of focused laser beam. While our model assume that all nanoparticles in the studied system are responsible for the heating process. Consequently, our calculations capture more collective heating effects. Second, the heat transfer equation is analytically solved in different ways: solving the thermal diffusion equation associated with continuity conditions of temperature and heat flux at interface in our work, and adopting the solution of Poisson equation in the previous study \cite{3}. Third, we take into account the influence of perfusion in the heating process. When the perfusion effects are ignored in Eq.(\ref{eq:3}) and (\ref{eq:4}), our approach gives $\Delta T \approx 3.04$, $6.08$ and 9.12 $^0C$ for $I_0 =2$, $4$, and $6$ $W/cm^2$, respectively. The perfusion term describes the energy removal due to microvascular network. The finding suggests that the perfusion effects on the heat transfer process play an important role.

Eq.(\ref{eq:5}) and (\ref{eq:6}) suggests that the steady-state temperature increase is linearly proportional to the excitation light intensity and nanoshell concentration. The finding is consistent with another theoretical model of photothermal effect \cite{2,3,33,34} and quantitatively agrees our experiment data. Several previous experimental studies \cite{3,4} have the same conclusions to confirm the validity of our model. 

\section{Conclusions}

In summary, we have proposed a new approach capturing effects of medium perfusion on thermal transport in biological systems for interpreting the laser-induced photothermal process and compared our theoretical analysis to experimental measurements. We synthesized gold nanoshell solution and injected it into chicken tissue samples before irradiating them by the NIR laser light. The temperature of samples as a function of time and the steady-state temperature trace versus the light intensity was measured and reported to compare to theoretical analysis. Our theoretical modeling successfully presents a quantitative understanding of the time dependent temperature and agrees with experiments. The influence of laser intensity and gold nanoshell concentrations on temperature rise due to laser illumination on chicken tissues has also been discussed. 

\begin{acknowledgments}
This work was funded by VAST: Young Researches project $N_0$ VAST.DLT.12/14-15. We would like to express gratitude to Jack Bernard for his comments. We would like to dedicate this work to the recently deceased Prof. Nguyen A. Viet who co-authored this paper.
\end{acknowledgments}

\end{document}